\documentclass{article}
\usepackage{spconf,amsmath,graphicx}
\usepackage{amssymb}
\usepackage{multirow}

\usepackage{flushend}

\usepackage{booktabs}
\usepackage{makecell}

\title{TG-Critic: A Timbre-Guided Model for Reference-Independent \\Singing Evaluation}
\name{Xiaoheng Sun $^{1,\emph{\text{*}}}$\thanks{\llap{\textsuperscript{*}}{These authors contributed equally to this research.}}, Yuejie Gao $^{1,\emph{\text{*}}}$, Hanyao Lin $^{2,\emph{\text{*}}}$ and Huaping Liu $^{1,\emph{\text{†}}}$\thanks{\llap{\textsuperscript{†}}{Corresponding author. E-mail: liuhuaping@corp.netease.com}}}
\address{
    $^1$ NetEase Cloud Music, Shanghai, China\\
    $^2$ Fudan University, Shanghai, China
    \vspace{-.35cm}
}

\begin{document}


	\ninept
	\maketitle
	\begin{abstract}
	
	Automatic singing evaluation independent of reference melody is a challenging task due to its subjective and multi-dimensional nature. As an essential attribute of singing voices, vocal timbre has a non-negligible effect and influence on human perception of singing quality. However, no research has been done to include timbre information explicitly in singing evaluation models. In this paper, a data-driven model TG-Critic is proposed to introduce timbre embeddings as one of the model inputs to guide the evaluation of singing quality. The trunk structure of TG-Critic is designed as a multi-scale network to summarize the contextual information from constant-Q transform features in a high-resolution way. Furthermore, an automatic annotation method is designed to construct a large three-class singing evaluation dataset with low human-effort. The experimental results show that the proposed model outperforms the existing state-of-the-art models in most cases.

	\end{abstract}
	\begin{keywords}
		Singing evaluation, timbre embedding, high-resolution network
	\end{keywords}

	\section{Introduction}
	Automatic singing evaluation aims to assess the quality of singing performances without the participation of music experts, thus reducing labor costs. The applications of the task include the distribution of singing content on the Internet and the discovery of musicians. With the popularity of online music services and karaoke singing applications, there is a growing demand for more advanced evaluation systems for singing voices, making automatic singing evaluation a hot research topic in recent years.
	
	Because singing evaluation by humans is empirically based and listener dependent, it is difficult to use automatic systems in substitution of human experts. However, previous studies have shown that humans rely more on some common, objective features than subjective preferences when judging the quality of singing \cite{nakano2006subjective}. These features such as intonation accuracy, tonal stability, and rhythm consistency can be detected automatically, which makes computational techniques feasible for this task. Also, the criteria of singing evaluation are inevitably multi-dimensional. A reliable singing evaluation system can be established only by observing objective metrics of a singing work in a similar way as music experts do.
	Depending on whether a reference melody is required, the existing automatic singing evaluation systems can be roughly divided into two types: reference-dependent approaches and reference-independent ones. A typical reference-dependent approach needs one ideal reference, such as an excellent singing recording or the correct melodic line. The more similar the singing is to the reference, the higher the score or the ranking level \cite{tsai2011automated}. Most studies focus on pitch accuracy \cite{lal2006comparison,cao2008study,gupta2018technical}, while some studies design hand-crafted audio features to obtain scores from other dimensions such as dynamic \cite{tsai2011automated, gupta2017perceptual}, rhythm \cite{tsai2011automated, gupta2017perceptual}, voice quality \cite{gupta2017perceptual}, timbre brightness \cite{omori1996singing, cao2008objective}, enthusiasm \cite{yu2015performance}, vibrato \cite{gupta2017perceptual}, tremolo \cite{polrolniczak2015computer}, \emph{etc.} There are also works using metric learning to map singing voices to embeddings for similarity comparison \cite{tan2019singing, zhang2021learn}. The advantages of reference-dependent approaches are high interpretability and low complexity, while the disadvantages are as below: \textbf{a)} Manual preparation/proofreading of references is needed; \textbf{b)} It encourages the imitation of references rather than personalized interpretation.
	
    For human experts, studies have shown that even if the melody is new to them, they can make a highly consistent evaluation \cite{nakano2006subjective}, indicating the feasibility of the reference-independent approaches. 
    The traditional reference-independent singing evaluation mainly focused on intonation characteristics \cite{nakano2006automatic, gupta2019automatic}. Since Zhang \emph{et al.} \cite{zhang2019automatic} proposed a convolutional neural network (CNN) based model trained on two-class data, the data-driven methods have been introduced into this field. Gupta \emph{et al.} \cite{gupta2020automatic} adopted a convolutional recurrent neural network (CRNN) framework, where the inputs are Mel-spectrograms along with pitch histograms. Huang \emph{et al.} \cite{huang2020spectral} further compared Mel-spectrogram with different mid-level input features such as constant-Q transform (CQT) and Chromagram. Using a similar CRNN structure, Li \emph{et al.} \cite{li2021training} trained multi-task models for both pitch and overall scores. Gupta \emph{et al.} \cite{gupta2021towards} transplanted the idea of pitch histogram to rhythm histogram for rhythm evaluation. The reference-independent methods are truly end-to-end, which helps to further reduce the need for human-effort and expand application scenarios.


	Although great performance has been achieved by previous studies, we notice that so far there is no such method that involves timbre information explicitly in the model. As previous studies \cite{cao2008study, cao2008objective, wapnick1997expert} have shown, the vocal timbre is an important attribute that affects human perception of singing.
    But unlike other attributes such as intonation and rhythm, timbre is a multifarious set of abstract sensory attributes \cite{mcadams2009perception} and is difficult to be extracted by models. 
	Therefore, we believe it is helpful to involve timbre information explicitly rather than to have the model learn the features from scratch.
	
	In this paper, we explore adding timbre embeddings as the model inputs and propose a \emph{timbre-guided} singing evaluation model named TG-Critic. 
	The trunk structure of TG-Critic is designed as a multi-scale CNN-based network to summarize the contextual information from the input CQT in a high-resolution way. In addition to the CQT, TG-Critic also takes a vector involving timbre information derived from the audio using a timbre embedding model. 
	Furthermore, to alleviate the problem of insufficient data, we propose an iterative automatic annotation method and construct a large singing dataset YJ-16K. 
    Experimental results show that by including timbre information explicitly, the model achieves higher accuracy and outperforms existing state-of-the-art methods.


	\begin{figure*}[t]
    \centering
    \includegraphics[width=16.5cm, trim={0.5cm 0.7cm 1.0cm 1cm}, clip]{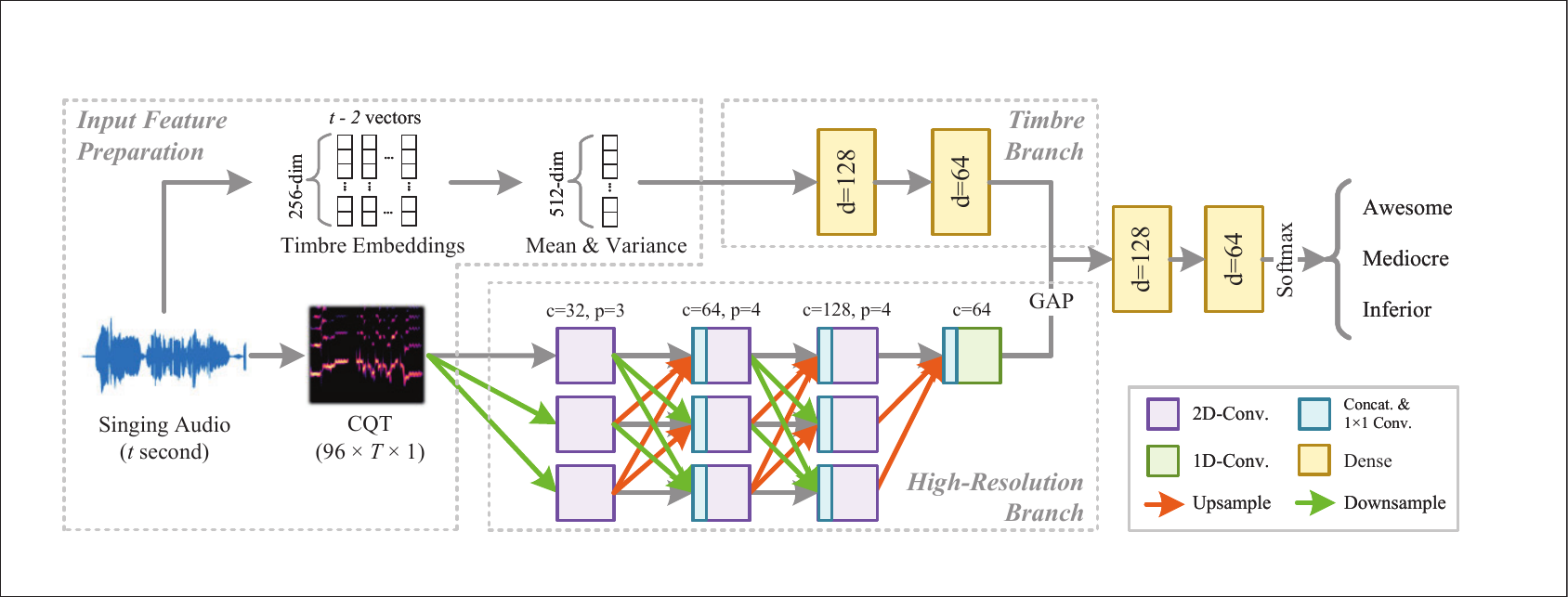}
    \caption{The overall architecture of TG-Critic. `c' denotes the channel number of the convolutional layer, while `p' denotes the pooling length along the frequency axis. All the 2D convolutional layers use the kernel size of $3\times3$, while the 1D convolutional layer's kernel length is 5. All the layers except the last dense layer are followed by the ELU activation function.}
    \label{fig:tg-critic}
    \end{figure*}

	\section{Approach}
	
	The overall architecture of TG-Critic is shown in Fig. \ref{fig:tg-critic}. It has two branches: 1) a \emph{High-Resolution Branch} mainly composed of convolutional layers processing the input CQT into high-level semantic representations for singing evaluation; 2) and a \emph{Timbre Branch} processing the timbre vector derived by a timbre embedding model. In this section, we first introduce the timbre embedding method and Timbre Branch. Then, the structure of High-Resolution Branch is detailed. Finally, we describe the construction of the singing dataset YJ-16K.
	

	\subsection{Timbre Embedding \& Timbre Branch}
	\label{sec:timbre}
	
	
	
	Timbre allows us to distinguish the difference between the two voices besides pitch, loudness, and duration \cite{M2019FMP}. To generate timbre representations that we can input into a deep learning model, we adopt the embedding model \emph{CROSS} proposed by Lee and Nam \cite{lee2019learning} and use the pretrained model\footnote{https://github.com/kyungyunlee/mono2mixed-singer} in experiments.  
	
	\emph{CROSS} is a metric learning-based embedding model designed for singer-relevant tasks, especially singer identification. A previous study has proved that singer identity mainly consists of two parts: timbre and singing style \cite{van2013analysis}. Therefore, the output vector should include enough information to distinguish vocal timbre and we believe it is suitable for timbre embedding. 
    
	The input of \emph{CROSS} is a three-second long audio clip while the output is an embedding vector of 256 dimensions. In experiments, we first slice each audio into segments of three seconds with the step size as one second and then derive the embedding vectors for each segment. Therefore, for audio with a length of $t$ seconds, a ``timbregram" with a size of $(t-2, 256)$ will be obtained, which depicts the variation of vocal timbre in the audio. To alleviate the influence of loudness, we perform min-max normalization on the timbregram. Finally, the mean and variance along the time axis are calculated and then concatenated as a vector of length 512, which we used as the input of TG-Critic. We excluded invalid timbre embeddings (singing voice frames are less than 70\%) in the calculation. 
	
	\begin{figure}[tb]
    \centering
    \fbox{\includegraphics[width=6.3cm, trim={4.9cm 1.4cm 20cm 1.3cm}, clip]{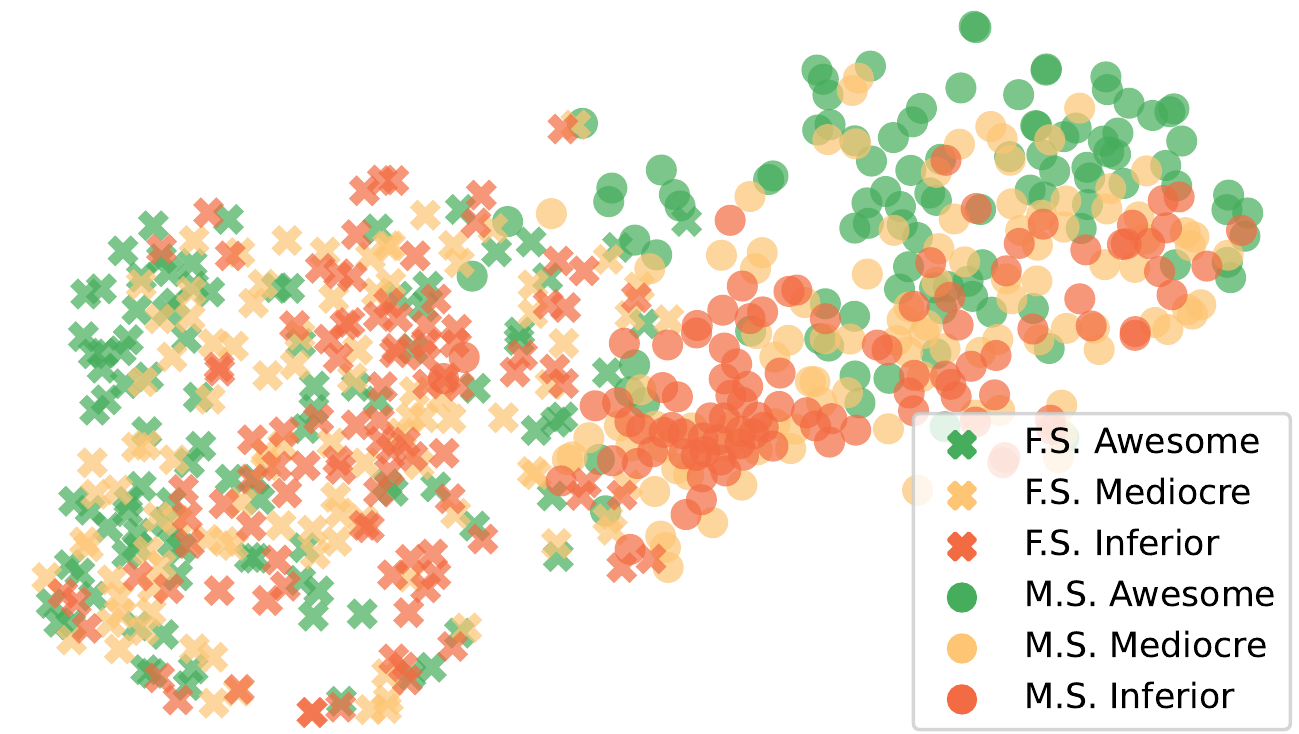}}
    \caption{The t-SNE visualization of the timbre vectors derived from singing samples with different qualities. M.S. and F.S. denote Male Singers and Female Singers respectively.} 
    \label{fig:tnse}
    \end{figure}
	
	To prove that information related to the singing quality is involved in the derived vectors, we randomly choose some singing samples with three-level quality annotations ``Awesome", ``Mediocre" and ``Inferior". Through the method mentioned above, from each audio, a 512-dimension timbre vector can be obtained. Then these timbre vectors are visualized using t-Distributed Stochastic Neighbor Embedding (t-SNE) \cite{van2008visualizing}, as shown in Fig. \ref{fig:tnse}. It shows that for male and female singing respectively, vectors of the same quality level are closer to each other than those of different levels. Furthermore, in an informal experiment, we use only these vectors to train a multilayer perceptron (MLP) classification model and get an accuracy of 62\%. These results indicate that the timbre vectors contain information helpful for automatic singing evaluation.
    
    As shown in Fig. \ref{fig:tg-critic}, Timbre Branch with two dense layers is designed to further process the 512-dimension timbre vectors into representations suitable for the fusion with the outputs of High-Resolution Branch. Finally, we obtain an output vector of 64 dimensions.
	
	

	\subsection{Structure of High-Resolution Branch}
	
	We use CQT as the input mid-level feature for High-Resolution Branch because previous studies \cite{huang2020spectral, li2021training, gupta2021towards} have proved its effectiveness. To extract features, the original audio data is first resampled to 16 kHz. For CQT computation, we set the hop size to 512. For each frame, there are 96 bins and each octave is covered by 24 bins. The minimum frequency is set to $E2$ to better match the pitch range of singing voices. For training and evaluation, we set the input length of CQT as 256 frames (roughly 8 seconds). Note that the input audio of TG-Critic can be of arbitrary length.
	
	For singing evaluation, long-term dependencies and local details are both indispensable. 
	As many studies \cite{zhang2019automatic, huang2020spectral, li2021training} did, we use a CNN-based structure as the backbone of our model to better detect local patterns. However, a few convolution layers cannot model long-distance dependence. On the other hand, the over-stacking of convolutional layers may lead to excessive parameters. Downsampling operations are usually introduced to expand the context range that convolution kernels can capture. But downsampling also causes a loss of detailed information. Besides, some studies \cite{huang2020spectral, li2021training} adopted the idea of CRNN, applying a recurrent layer after CNN to capture long-term dependencies. However, the involvement of RNN undoubtedly reduces the model's efficiency.
	
	Inspired by the success of HRNet \cite{wang2020deep} designed for computer vision tasks, we introduce a similar multi-scale structure into the proposed model. In such a structure, downsampling is performed on the input features to obtain low-resolution representations to facilitate the extraction of global information. After the process of convolutional layers, the low-resolution features are upsampled and merged with high-resolution features. The procedure of downsampling and upsampling is repeated several times, and it helps aggregate contextual information from different scales. Because of the existence of high-resolution features, the detailed patterns are not degraded during the process. 
    Specifically, the input CQT of $T$ frames is first downsampled to $T/2$ and $T/4$ over the time axis (we assume that $T$ is an integral multiple of 4). These features are then processed by $3\times3$ convolutional layers. After average pooling over the frequency axis to summarize frequency bands, the features are rescaled and fused. Average pooling and nearest neighbor interpolation are used to perform rescaling. For the fusion operation, the concatenation is followed by a $1\times1$ convolutional layer to exchange information and adjust the channel number. Meanwhile, on the frequency axis the features are gradually downsampled. After three multi-scale processes, the features are fused again. We reshape the obtained feature to $(T, 768)$ and use a 1D convolutional layer to adjust the channel numbers to 64. After global average pooling, a vector of 64 dimensions is obtained.
	
	
	After obtaining the output vectors from the two branches, we concatenate them and send the fused vector into a dense layer with exponential linear unit (ELU) activation \cite{clevert2015fast} to exchange information. Finally, we apply a dense layer with softmax activation to produce the classification result for singing quality.
	

	\subsection{Construction of YJ-16K Dataset}
	
	\begin{figure}[t]
    \centering
    \includegraphics[width=8cm, trim={2.5cm 5cm 2.5cm 3.7cm}, clip]{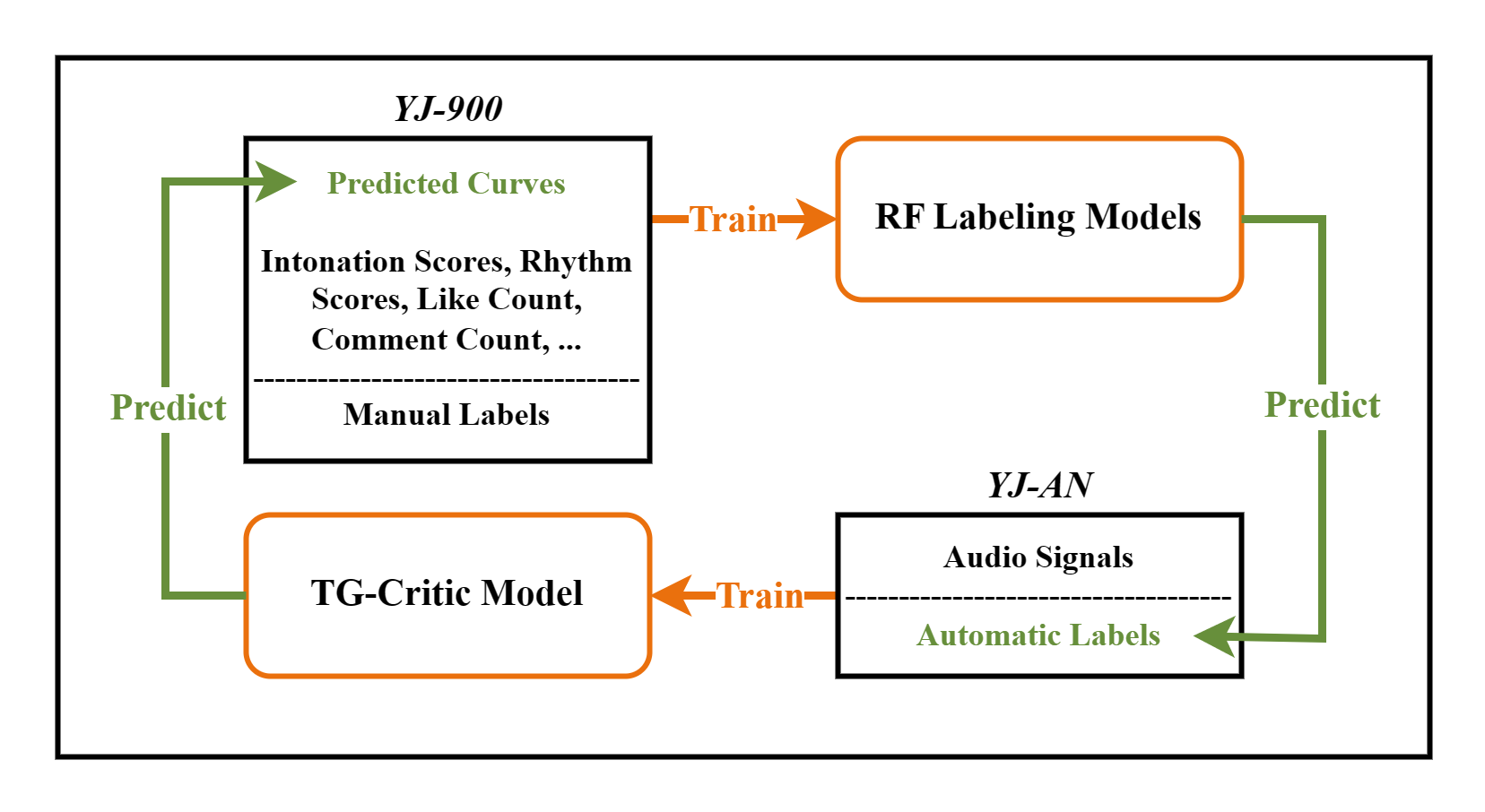}
    \caption{The iterative automatic annotation of YJ-16K dataset.}
    \label{fig:annotation}
    \end{figure}

	We construct a three-class singing dataset named YJ-16K. The dataset contains 32,623 users' unaccompanied singing pieces from YinJie\footnote{https://k.163.com/ is a karaoke application.}. We collect metadata from the application including scoring data such as intonation scores, rhythm scores, \emph{etc.}, and user behavior data such as the numbers of likes and comments for each song. The dataset contains a variety of singing from poor to professional. We aim to label these samples into three classes: Awesome (\textbf{A}) - works with excellent execution and impressive voice quality; Mediocre (\textbf{M}) - works with good intonation, but with ordinary expressiveness; Inferior (\textbf{I}) - works with weak intonation or poor expressiveness.

    To reduce labor costs of labeling, we train models for automatic annotation. First, a subset of 894 samples named YJ-900 is separated and manually annotated by music experts. Using YJ-900's metadata and manual labels, we train two random forest (RF) binary classification models to separate Awesome and Inferior samples respectively. Then, the RF models are used to annotate the rest of the original dataset, called YJ-AN. However, these automatic labels are not reliable enough, because the metadata we obtained was affected by various external factors. For example, covers of hit songs often tend to get higher user ratings.
    
    To improve the reliability of annotations, we design an ``iterative automatic annotation" method as shown in Fig. \ref{fig:annotation}. We first train the TG-Critic model using the audio signals and automatic labels of YJ-AN. By this model, a predicted ``evaluation curve" is obtained for each audio sample of YJ-900 (as TG-Critic's inputs are 8-second segments), which is then used as a part of the inputs of the RF Models in the next iteration. The iteration repeates four times, and the detail of the finally obtained dataset is shown in Table \ref{tab:YJ-16K}. The effects of iterative automatic annotation can be seen in Section \ref{ssc:effect_of_IA}.

	\begin{table}[!tb]
	\setlength{\tabcolsep}{1.8mm}{
	    \centering
	    \begin{tabular}{c|c|c|ccc|cc}
	    \toprule[1.2pt]
	         \multicolumn{2}{c|}{YJ-16K} & Total & \textbf{A} & \textbf{M} & \textbf{I} & M.S. & F.S. \\ \hline\hline
	         YJ-900 & test & 894 & 296 & 302 & 296 & 430 & 464 \\ \hline
	         \multirow{3}{*}{YJ-AN} & total & 15000 & 5000 & 5000 & 5000 & 7683 & 7317 \\ 
	         & train & 14100 & 4700 & 4700 & 4700 & 7203 & 6897 \\ 
	         & valid & 900 & 300 & 300 & 300 & 480 & 420 \\ 
	    \bottomrule[1.2pt]
        \end{tabular}
        \caption{The number of samples in YJ-16K dataset. M.S. denotes Male Singers while F.S. denotes Female Singers. }
        \label{tab:YJ-16K}
    }
	\end{table}

	

	\section{Evaluation}

	\subsection{Ablation Studies}
	\label{sec:ablation}
	
	To demonstrate how much the timbre vectors contribute to the model, we compare the performance of the models with and without Timbre Branch. For the structure with Timbre Branch (the proposed TG-Critic), two models are trained by different training strategies: \textbf{a)} \emph{TG-Critic-1S}: The High-Resolution Branch and the Timbre-Branch are trained together in one step; \textbf{b)} \emph{TG-Critic-2S}: A two-step training strategy is applied where the High-Resolution Branch is first trained and frozen, and then the Timbre Branch is trained using smaller learning rate and fewer epochs. We also trained a model \emph{CQT-Only} with only High-Resolution Branch and CQT input for comparison. Among the models mentioned above, \emph{TG-Critic-1S} and \emph{CQT-Only} are both trained by 200 epochs with a learning rate of 0.0001. \emph{TG-Critic-2S} is trained 100 epochs for each step, and the learning rate is set as 0.0001 in the first step and 0.00005 in the second step. For all three models, the Adam optimizer is applied in training to minimize categorical cross-entropy, and one epoch consists of 500 batches where the batch size is set as 32.
	
	\begin{table}[tb]
	\setlength{\tabcolsep}{1.8mm}{
	    \centering
	    \begin{tabular}{c|ccc|ccc|c}
	    \toprule[1.2pt]
	         \multirow{2}{*}{Model} & \multicolumn{3}{c|}{Precision (\%)} & \multicolumn{3}{c|}{Recall (\%)} & \multirow{2}{*}{\makecell[c]{Acc.\\(\%)}} \\ \cline{2-7}
	         & \textbf{A} & \textbf{M} & \textbf{I} & \textbf{A} & \textbf{M} & \textbf{I} & \\ 
	         \hline \hline

	        \emph{TG-Critic-1S}  & 83.5 & 71.6 & 84.0 & 90.5 & 69.2 & 79.7 & 79.8 \\ 
	        \emph{TG-Critic-2S}  & \textbf{87.2} & \textbf{73.6} & 86.7 & 89.9 & \textbf{75.5} & 81.8 & \textbf{82.3} \\ 
	        \emph{CQT-Only}    & 84.3 & 69.8 & 79.5 & 88.9 & 63.6 & \textbf{82.4} & 78.2 \\ 
	        \emph{TG-Simple}     & 82.1 & 68.4 & \textbf{88.7} & \textbf{92.9} & 72.5 & 71.6 & 79.0 \\
	       
	    \bottomrule[1.2pt]
        \end{tabular}
        \caption{Results of ablation studies on YJ-900 dataset. }
        \label{tab:ablation}
	}
	\end{table}
	
	As shown in Table \ref{tab:ablation}, the model without Timbre Branch performs worse than the models with Timbre Branch. Among the three classes, \emph{CQT-Only} shows the highest recall and the lowest precision on the inferior singings, indicating the model without explicit timbre information is inclined to underestimate the singing quality. For the two models with Timbre Branch, \emph{TG-Critic-2S} using the two-step training strategy presents a significant enhancement in performance. We believe this is mainly because Timbre Branch is a much simpler structure than High-Resolution Branch, and it is difficult for the two branches to converge simultaneously in one-step training. 
	
    Furthermore, to evaluate the idea of using a ``high-resolution" network, we remove the High-Resolution Branch in the model named \emph{TG-Simple}. Instead, a simple CNN structure stacked with three $3\times3$ convolutional layers is used. We carefully adjust the filter numbers in the layers to remain a similar number of parameters, and the same training strategy as \emph{TG-Critic-2S} is used to ensure the model's best performance. As shown in Table \ref{tab:ablation}, the accuracy decreases by 3.3\% compared to \emph{TG-Critic-2S}, proving that the high-resolution structure helps to produce better representations for singing evaluation. 
	

	\subsection{Comparison with Previous Works}
	
	\begin{table}[tb]
	\setlength{\tabcolsep}{1.7mm}{
	    \centering
	    \begin{tabular}{c|c|c|cc|cc}
	    \toprule[1.2pt]
	         \multirow{2}{*}{Model} & \multirow{2}{*}{Param.} & YJ-900 & \multicolumn{2}{c|}{PESnQ-DS} & \multicolumn{2}{c}{NUS48E} \\ \cline{3-7}
	         & & Acc. & Acc. & Corr. & Acc. & Corr. \\
	         \hline\hline
	         \emph{Kuaishou} \cite{zhang2019automatic} & 1.97M & 68.3 & 85.0 & 0.858 & 68.8 & 0.497 \\
	         \emph{NUS20} \cite{huang2020spectral}     & 0.72M & 76.3 & 85.0 & 0.930 & 68.8 & 0.552 \\
	         \emph{NUS21} \cite{li2021training}        & 1.45M & 78.4 & 85.0 & 0.925 & 72.9 & 0.548 \\
	         \hline
	         \emph{TG-Critic-1S}  & \multirow{2}{*}{0.82M} & 79.8 & 80.0 & 0.927 & 72.9 & \textbf{0.671} \\
	         \emph{TG-Critic-2S}  & & \textbf{82.3} & \textbf{95.0} & \textbf{0.933} & \textbf{77.1} & 0.631 \\
	       
	    \bottomrule[1.2pt]
        \end{tabular}
        \caption{Results of the proposed and baseline models on YJ-900, PESnQ-DS, NUS48E datasets. The accuracy values are percentiles. (Acc. = Accuracy, Corr. = Pearson Correlation Coefficient.)}
        \label{tab:comparison}
	}
	\end{table}

	To compare with previous works, we reproduce three baseline models: \emph{Kuaishou} \cite{zhang2019automatic} (Bi-DenseNet-27 model), \emph{NUS20} \cite{huang2020spectral} (CPH-CRNN model), and \emph{NUS21} \cite{li2021training} (Framework 3). All three models are reference-independent singing evaluation models based on deep learning. 
	We follow the hyperparameters described in the papers to ensure the models' performances on our training dataset YJ-AN. 
	
    In addition to YJ-900, we use two public datasets PESnQ-DS \cite{gupta2017perceptual} and NUS48E \cite{duan2013nus} for tests. Recordings of PESnQ-DS (two songs, each performed by 10 singers) cover singing ability from poor to professional and they are annotated with overall scores, pitch scores, and rhythm scores. NUS48E contains 48 samples (20 songs) performed by 12 singers between mediocre and good level, without quality annotations. Besides, three of the PESnQ-DS recordings are from the NUS48E dataset.
    
    To collect subjective ratings on NUS48E, we recruit 12 human judges to give an overall singing quality score out of 5 to each recording compared to the reference singing samples. We also conduct the same rating procedure on PESnQ-DS and find our scores are highly consistent with the overall scores provided by the authors, achieving the Pearson correlation coefficient (PCC) of 0.91, which proves the reliability of our annotations.
    
    Considering the singers' singing ability of NUS48E is from mediocre to good level, we divide the NUS48E samples into classes \textbf{A} and \textbf{M} using a threshold (resulting in 29 as \textbf{A} and 19 as \textbf{M}). On PESnQ-DS, we calculate the average scores of all annotators (12 by us and 5 by \cite{gupta2017perceptual}) for comparison, and divide samples into classes \textbf{A}, \textbf{M}, and \textbf{I} by two thresholds (resulting in 8 as \textbf{A}, 7 as \textbf{M}, and 5 as \textbf{I}). 

    The experimental results are shown in Table \ref{tab:comparison}. Compared with other models, \emph{TG-Critic-2S} achieves the highest accuracy for all three datasets. To make a comprehensive comparison, we obtain a weighted score for each prediction using the output probability distribution. We annotate predicted probabilities for class \textbf{A}, \textbf{M}, and \textbf{I} be $P_A$, $P_M$, and $P_I$ respectively. Then the weighted score is calculated as: $P_A\times1.0+P_M\times0.5+P_I\times0.0$, ranging within $[0, 1]$. The PCCs between the weighted scores and the ground truth are shown in Table \ref{tab:comparison}. The high correlations indicate that these scores from probability distribution are capable of measuring the singing quality. \emph{TG-Critic-1S} and \emph{TG-Critic-2S} show better performance than the baseline models, clearly confirming the effectiveness and robustness of our proposed model\footnote{The detail annotations for PESnQ-DS and NUS48E and further experimental results are available at https://github.com/YuejieGao/TG-CRITIC\label{fn:github}}. 
	
	
    \subsection{Effects of Iterative Automatic Annotation}\label{ssc:effect_of_IA}
	
	\begin{figure}[tb]
    \centering
    \includegraphics[width=6.3cm, trim={3cm 4.6cm 2cm 4.5cm}, clip]{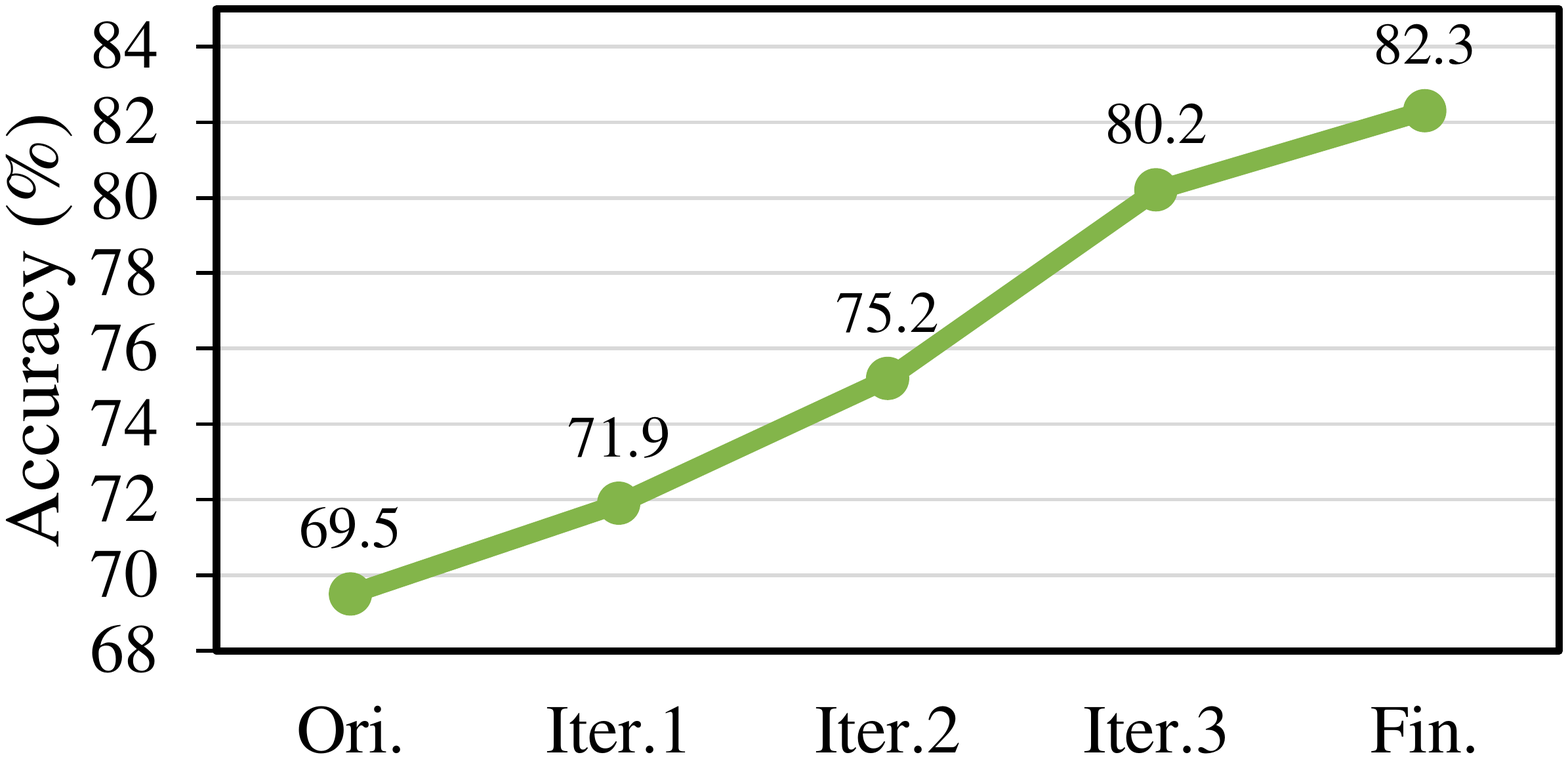}
    \caption{Performance of models trained on datasets of different annotation iterations.} 
    \label{fig:iteration}
    \end{figure}
    
    
    To demonstrate the effects of iterative automatic annotation, we train models with datasets after each annotation iteration using the two-step training strategy mentioned in Section \ref{sec:ablation}. Then, the models are evaluated on the manually annotated YJ-900. 
    
    As shown in  Fig. \ref{fig:iteration}, the model's performance improves significantly after each iteration, especially for the first three ones, which indicates that the annotations are increasingly closer to human assessment. The results prove that the iterative annotation process is effective in producing reliable annotations.
    
	
	\section{Conclusion}
	In this paper, we have proposed TG-Critic, a timbre-guided singing evaluation model independent of the reference melody. The proposed model includes timbre information explicitly by using timbre embedding as one of the model inputs. A multi-scale structure is introduced to process the CQT features in a high-resolution way. We also construct a large singing dataset YJ-16K with annotations labeled by an iterative automatic annotation method. Experimental results show the proposed model outperforms the existing state-of-the-art models in most cases.
	
	
	\bibliographystyle{IEEEbib}
	\bibliography{strings,refs}

\begin{thebibliography}{10}

\bibitem{nakano2006subjective}
Tomoyasu Nakano, Masataka Goto, and Yuzuru Hiraga,
\newblock ``Subjective evaluation of common singing skills using the rank
  ordering method,''
\newblock in {\em Ninth International Conference on Music Perception and
  Cognition}. Citeseer, 2006, pp. 1507--1512.

\bibitem{tsai2011automated}
Wei-Ho Tsai and Hsin-Chieh Lee,
\newblock ``An automated singing evaluation method for karaoke systems,''
\newblock in {\em 2011 IEEE International Conference on Acoustics, Speech and
  Signal Processing (ICASSP)}. IEEE, 2011, pp. 2428--2431.

\bibitem{lal2006comparison}
Partha Lal,
\newblock ``A comparison of singing evaluation algorithms,''
\newblock in {\em Proceedings of INTERSPEECH 2006}, 2006, vol.~5, pp.
  2298--2301.

\bibitem{cao2008study}
Chuan Cao, Ming Li, Jian Liu, and Yonghong Yan,
\newblock ``A study on singing performance evaluation criteria for untrained
  singers,''
\newblock in {\em 2008 9th International Conference on Signal Processing}.
  IEEE, 2008, pp. 1475--1478.

\bibitem{gupta2018technical}
Chitralekha Gupta, Haizhou Li, and Ye~Wang,
\newblock ``A technical framework for automatic perceptual evaluation of
  singing quality,''
\newblock {\em APSIPA Transactions on Signal and Information Processing}, vol.
  7, 2018.

\bibitem{gupta2017perceptual}
Chitralekha Gupta, Haizhou Li, and Ye~Wang,
\newblock ``Perceptual evaluation of singing quality,''
\newblock in {\em 2017 Asia-Pacific Signal and Information Processing
  Association Annual Summit and Conference (APSIPA ASC)}. IEEE, 2017, pp.
  577--586.

\bibitem{omori1996singing}
Koichi Omori, Ashutosh Kacker, Linda~M Carroll, William~D Riley, and Stanley~M
  Blaugrund,
\newblock ``Singing power ratio: quantitative evaluation of singing voice
  quality,''
\newblock {\em Journal of Voice}, vol. 10, no. 3, pp. 228--235, 1996.

\bibitem{cao2008objective}
Chuan Cao, Ming Li, Jian Liu, and Yonghong Yan,
\newblock ``An objective singing evaluation approach by relating acoustic
  measurements to perceptual ratings,''
\newblock in {\em Proceedings of INTERSPEECH 2008}, 2008, pp. 2058--2061.

\bibitem{yu2015performance}
Yang Yu, Weisi Lin, Dong-Yan Huang, Minghui Dong, and Haizhou Li,
\newblock ``Performance scoring of singing voice,''
\newblock in {\em 2015 International Conference on Asian Language Processing
  (IALP)}. IEEE, 2015, pp. 119--122.

\bibitem{polrolniczak2015computer}
Edward P{\'o}{\l}rolniczak and Micha{\l} Kramarczyk,
\newblock ``Computer assessment of tremolo feature in the context of evaluation
  of singing quality,''
\newblock in {\em 2015 Signal Processing: Algorithms, Architectures,
  Arrangements, and Applications (SPA)}. IEEE, 2015, pp. 157--161.

\bibitem{tan2019singing}
Terry Tan,
\newblock ``Singing evaluation based on deep metric learning,''
\newblock in {\em Proceedings of the 2019 3rd International Symposium on
  Computer Science and Intelligent Control}, 2019, pp. 1--5.

\bibitem{zhang2021learn}
Huan Zhang, Yiliang Jiang, Tao Jiang, and Hu~Peng,
\newblock ``Learn by referencing: Towards deep metric learning for singing
  assessment.,''
\newblock in {\em the 22nd International Society for Music Information
  Retrieval Conference (ISMIR)}, 2021, pp. 825--832.

\bibitem{nakano2006automatic}
Tomoyasu Nakano, Masataka Goto, and Yuzuru Hiraga,
\newblock ``An automatic singing skill evaluation method for unknown melodies
  using pitch interval accuracy and vibrato features,''
\newblock in {\em Proceedings of INTERSPEECH 2006}, 2006, pp. 1706--1709.

\bibitem{gupta2019automatic}
Chitralekha Gupta, Haizhou Li, and Ye~Wang,
\newblock ``Automatic leaderboard: Evaluation of singing quality without a
  standard reference,''
\newblock {\em IEEE/ACM Transactions on Audio, Speech, and Language
  Processing}, vol. 28, pp. 13--26, 2019.

\bibitem{zhang2019automatic}
Ning Zhang, Tao Jiang, Feng Deng, and Yan Li,
\newblock ``Automatic singing evaluation without reference melody using
  bi-dense neural network,''
\newblock in {\em 2019 IEEE International Conference on Acoustics, Speech and
  Signal Processing (ICASSP)}. IEEE, 2019, pp. 466--470.

\bibitem{gupta2020automatic}
Chitralekha Gupta, Lin Huang, and Haizhou Li,
\newblock ``Automatic rank-ordering of singing vocals with twin-neural
  network.,''
\newblock in {\em the 21st International Society for Music Information
  Retrieval Conference (ISMIR)}, 2020, pp. 416--423.

\bibitem{huang2020spectral}
Lin Huang, Chitralekha Gupta, and Haizhou Li,
\newblock ``Spectral features and pitch histogram for automatic singing quality
  evaluation with crnn,''
\newblock in {\em 2020 Asia-Pacific Signal and Information Processing
  Association Annual Summit and Conference (APSIPA ASC)}. IEEE, 2020, pp.
  492--499.

\bibitem{li2021training}
Jinhu Li, Chitralekha Gupta, and Haizhou Li,
\newblock ``Training explainable singing quality assessment network with
  augmented data,''
\newblock in {\em 2021 Asia-Pacific Signal and Information Processing
  Association Annual Summit and Conference (APSIPA ASC)}. IEEE, 2021, pp.
  904--911.

\bibitem{gupta2021towards}
Chitralekha Gupta, Jinhu Li, and Haizhou Li,
\newblock ``Towards reference-independent rhythm assessment of solo singing,''
\newblock in {\em 2021 Asia-Pacific Signal and Information Processing
  Association Annual Summit and Conference (APSIPA ASC)}. IEEE, 2021, pp.
  912--919.

\bibitem{wapnick1997expert}
Joel Wapnick and Elizabeth Ekholm,
\newblock ``Expert consensus in solo voice performance evaluation,''
\newblock {\em Journal of Voice}, vol. 11, no. 4, pp. 429--436, 1997.

\bibitem{mcadams2009perception}
Stephen McAdams and Bruno~L Giordano,
\newblock ``The perception of musical timbre,''
\newblock {\em The Oxford handbook of music psychology}, pp. 72--80, 2009.

\bibitem{M2019FMP}
Müller Meinard and Frank Zalkow,
\newblock ``Fmp notebooks: Educational material for teaching and learning
  fundamentals of music processing.,''
\newblock in {\em the 20th International Society for Music Information
  Retrieval Conference (ISMIR)}, 2019, pp. 573--580.

\bibitem{lee2019learning}
Kyungyun Lee and Juhan Nam,
\newblock ``Learning a joint embedding space of monophonic and mixed music
  signals for singing voice,''
\newblock in {\em the 20th International Society for Music Information
  Retrieval Conference (ISMIR)}, 2019, pp. 295--302.

\bibitem{van2013analysis}
JMH van Balen, John~Ashley Burgoyne, Frans Wiering, and Remco~C Veltkamp,
\newblock ``An analysis of chorus features in popular song,''
\newblock in {\em the 14th International Society of Music Information Retrieval
  Conference (ISMIR)}, 2013, pp. 107--112.

\bibitem{van2008visualizing}
Laurens Van~der Maaten and Geoffrey Hinton,
\newblock ``Visualizing data using t-sne.,''
\newblock {\em Journal of machine learning research}, vol. 9, no. 11, 2008.

\bibitem{wang2020deep}
Jingdong Wang, Ke~Sun, Tianheng Cheng, Borui Jiang, Chaorui Deng, Yang Zhao,
  Dong Liu, Yadong Mu, et~al.,
\newblock ``Deep high-resolution representation learning for visual
  recognition,''
\newblock {\em IEEE transactions on pattern analysis and machine intelligence},
  vol. 43, no. 10, pp. 3349--3364, 2020.

\bibitem{clevert2015fast}
Djork-Arn{\'e} Clevert, Thomas Unterthiner, and Sepp Hochreiter,
\newblock ``Fast and accurate deep network learning by exponential linear units
  (elus),''
\newblock {\em arXiv preprint arXiv:1511.07289}, 2015.

\bibitem{duan2013nus}
Zhiyan Duan, Haotian Fang, Bo~Li, Khe~Chai Sim, and Ye~Wang,
\newblock ``The nus sung and spoken lyrics corpus: A quantitative comparison of
  singing and speech,''
\newblock in {\em 2013 Asia-Pacific Signal and Information Processing
  Association Annual Summit and Conference (APSIPA ASC)}. IEEE, 2013, pp. 1--9.

\end{thebibliography}
\end{document}